\documentclass[preprint, aps, pra]{revtex4-1}

\usepackage[margin=1in]{geometry}                  
\usepackage{graphicx}
\usepackage{bm}
\usepackage{amsmath,amssymb,amsfonts}
\usepackage{color}
\usepackage[hidelinks]{hyperref}
\usepackage{color}
\definecolor{dkgreen}{rgb}{0,0.6,0}
\definecolor{gray}{rgb}{0.5,0.5,0.5}

\hypersetup{
    colorlinks,%
    citecolor=blue,%
    filecolor=blue,%
    linkcolor=blue,%
    urlcolor=blue
}

\definecolor{orange}{rgb}{1,0.5,0}

\begin{document}

\title{Velocity-selective EIT measurement of potassium Rydberg states}
\author{Wenchao Xu}
\affiliation{Department of Physics,  University of Illinois at Urbana-Champaign, Urbana, Illinois 61801, USA}
\author{Brian DeMarco}
\affiliation{Department of Physics,  University of Illinois at Urbana-Champaign, Urbana, Illinois 61801, USA}
\email{bdemarco@illinois.edu}

\date{\today}

\begin{abstract}
We demonstrate a velocity selection scheme that mitigates suppression of electromagnetically induced transparency (EIT) by Doppler shifts for low--high EIT probe--coupling wavelength ordering.  An optical pumping beam counter-propagating with the EIT probe beam transfers atoms between hyperfine states in a velocity selective fashion.  Measurement of the transmitted probe beam synchronous with chopping of the optical pumping beam enables a Doppler-free EIT signal to be detected. Transition frequencies between 5P$_{1/2}$ and $n$S$_{1/2}$ states for $n=$26, 27, and 28 in $^{39}$K are obtained via EIT spectroscopy in a heated vapor cell with a probe beam stabilized to the 4S$_{1/2}\rightarrow$5P$_{1/2}$ transition.  Using previous high-resolution measurements of the 4S$_{1/2}\rightarrow$nS$_{1/2}$ transitions, we make a determination of the absolute frequency of the 4S$_{1/2}\rightarrow$5P$_{1/2}$ transition.  Our measurement is shifted by 560 MHz from the currently accepted value with a two-fold improvement in uncertainty.  These measurements will enable novel experiments with Rydberg-dressed ultracold Fermi gases composed of $^{40}$K atoms.
\end{abstract}
\maketitle




Introducing long-range interactions between atoms trapped in optical lattices has attracted intense interest because of the possibility to realize novel quantum phases of matter, such as quantum magnetism \cite{Gorshkov2011,moses2015} and topological superfluids \cite{Cooper2009}. Several experimental approaches have been pursued, including trapping polar molecules \cite{Yan2013} and atomic species with large magnetic dipole moments \cite{Lu2012,stuhler2005,aikawa2012}. Rydberg atoms, which have a valence electron excited to a state with high principle quantum number $n$ and therefore exhibit strong van der Waals inter-atomic interactions, are also a potential candidate. Interaction-induced blockades \cite{Viteau2011} and self-assembled crystals of Rydberg atoms have been observed experimentally in optical lattices \cite{zeiher2015,Schau2012,schauss2015}. However, the typical lifetime of Rydberg states is too short for a quantum many-body system to equilibrate, which hinders exploring exotic quantum phases. Recent theoretical proposals have suggested Rydberg-dressed states as a technique to avoid this problem.  In this method, a small and adjustable fraction of a Rydberg state is coherently mixed into the ground state wavefunction to extend the lifetime and tune the interaction length scale to be comparable to the lattice spacing \cite{Macr2014,johnson2010}.





Experimental work on ultracold Rydberg gases has largely focused on bosonic $^{87}$Rb atoms \cite{saffman2010}.  Rydberg dressing, for example, was investigated using trapped $^{87}$Rb atoms without a lattice \cite{1367-2630-16-6-063012}.  However, the Rydberg-dressed interaction could not be detected.  Realizing Rydberg-dressed interactions between fermionic atoms trapped in an optical lattice would be an exciting step toward resolving puzzles related to strongly correlated electronic solids \cite{mattioli2013,li2015}.  $^{40}$K atoms are an attractive candidate for Rydberg dressing of a fermionic species because they are readily trapped in a lattice in Mott insulator and band insulator states \cite{jordens2008mott,schneider2008metallic}.  Furthermore, Rydberg states can be accessed using diode lasers via the $4S\rightarrow5P$ and $5P\rightarrow nS$ transitions at 404.8 and 1003.7--977.3 nm for $n=$20--40, respectively.  Also, using the $5P$ state as an intermediate level enhances the Rydberg dressed lifetime compared with the D2 transition employed in experiments with Rb so far, since the natural lifetime is approximately six times longer.  While the absolute frequencies of the $4S\rightarrow nS$ transitions are known with approximately $\pm$10~MHz uncertainty \cite{lorenzen1981}, the $4S\rightarrow5P$ frequency has a relatively large reported $\pm150$~MHz uncertainty \cite{johansson1972} and, to our knowledge, a measurement of the absolute frequency has not been carried out. Therefore, determining the $4S\rightarrow5P$ wavelength and developing a method for spectroscopically resolving this transition is a necessary first step toward Rydberg dressing with $^{40}$K.

In this paper, we describe using electromagnetically induced transparency (EIT) \cite{Fleischhauer2005} to probe Rydberg transitions for $^{39}$K atoms contained in a heated vapor cell. The primary advantage of detecting Rydberg levels in $^{39}$K is its high natural abundance compared with $^{40}$K.  The $235\pm2$~MHz isotope shift for the $4S_{1/2}\rightarrow5P_{1/2}$ transition is known with high accuracy \cite{behrle2011}, and the approximately 100 MHz isotope shifts for the $nS$ states with $n>9$ are consistent with the Bohr mass shift within 1~MHz \cite{niemax1980,pendrill1982}.  Therefore, once the transitions are identified using $^{39}$K, the laser frequencies can be straightforwardly shifted using acousto-optic modulators to address $^{40}$K.   While using a vapor cell is simpler and less resource intensive than a cold gas, the Doppler effect suppresses the EIT signal for the high--low frequency ordering that we employ \cite{boon1999,urvoy2013}.  To overcome this effect, we have developed a strategy for velocity-selective EIT spectroscopy and detected EIT for the 404.8~nm probe transition induced by a coupling laser tuned to $5P\rightarrow nS$ for $n=26$--28.

\begin{figure}[!htp]
\includegraphics[width=0.8\columnwidth]{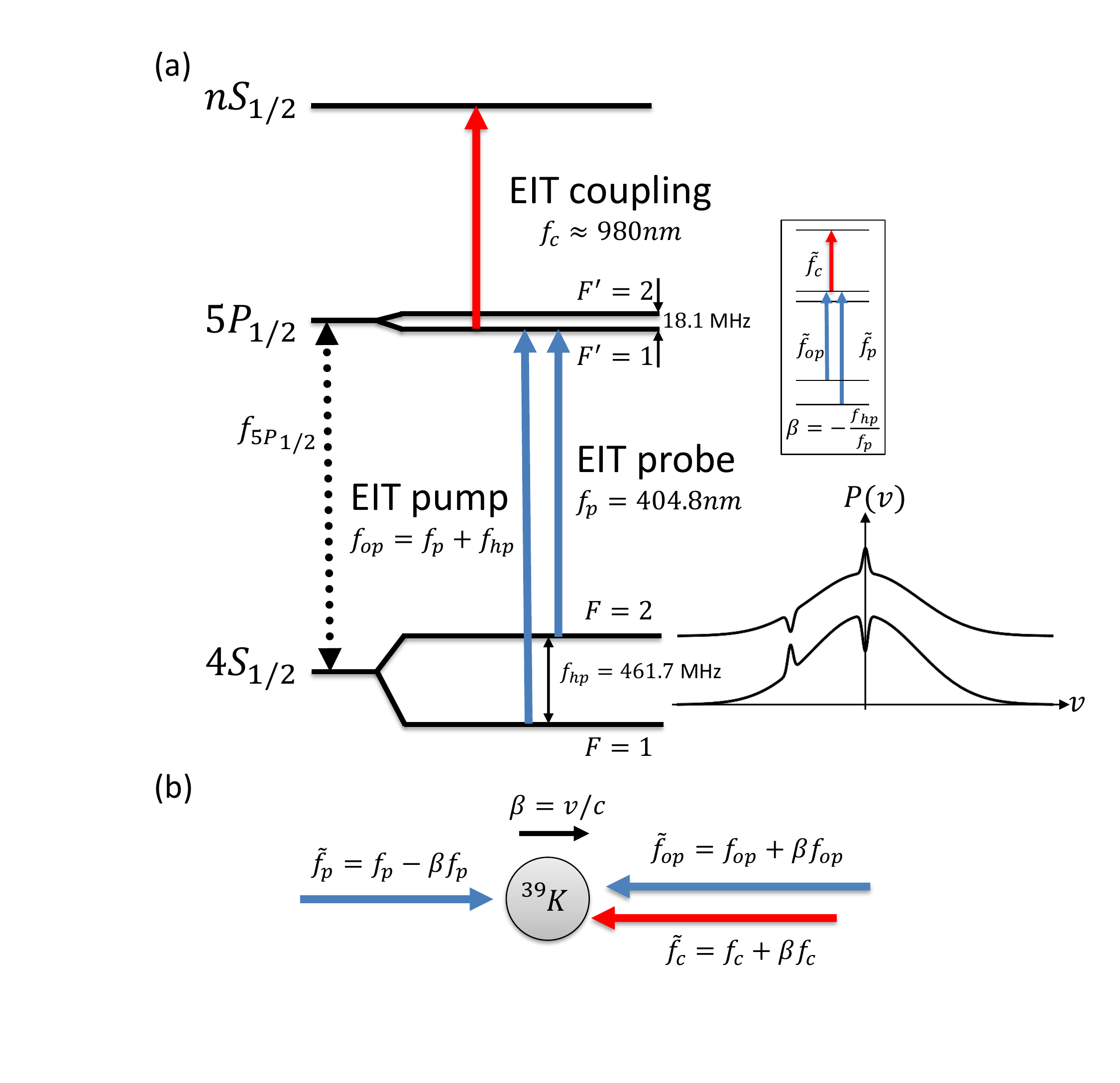}
\caption{(a): Energy-level diagram of $^{39}$K and corresponding transition wavelengths. The three-level ladder system used for EIT includes ground state $4S_{1/2}$, intermediate state $5P_{1/2}$, and Rydberg state $nS_{1/2}$. The optical pumping beam transfers atoms between the $F=1$ and $F=2$ ground states in a velocity sensitive fashion, as shown in the schematic Gaussian velocity distributions $P(v)$.  The populations imbalances at $v=0$ and $v=-c f_{hp}/f_p$ are modulated at the chopping frequency of the optical pumping beam. (b) Frequencies of the pump beam $\tilde{f}_{op}$, probe beam $\tilde{f}_{p}$, and coupling beam $\tilde{f}_c$ in the atomic rest frame, where $\beta=v/c$ and $v$ is atomic velocity. The inset shows the frequencies in the atomic rest frame for velocity $v=-c f_{hp}/f_p$, in which case the frequency of EIT pump and probe beam are exchanged.}
\label{fig:EnergyLevels}
\end{figure}

The energy level diagram relevant to our EIT measurement is shown in Fig.~\ref{fig:EnergyLevels}(a). This three-level ladder system includes ground state $4S_{1/2}$, intermediate state $5P_{1/2}$, and highly excited Rydberg state $nS_{1/2}$. The hyperfine splitting $f_{hp}$ between the $F=2$ and $F=1$ (where $F$ is the quantum number for the total angular momentum) ground states is 461.7~MHz \cite{arimondo1977}, and they are equally populated at room temperature. A weak probe beam with wavelength $\lambda_p=404.8$~nm and frequency $f_p$ is locked to the $4S\rightarrow 5P_{1/2}$ transition via frequency modulation spectroscopy. The 18.1 MHz hyperfine splitting for the excited $5P_{1/2}$ state \cite{arimondo1977} is not resolved by the probe laser. EIT for the $4S\rightarrow 5P_{1/2}$ transition is induced by an intense coupling beam with frequency $f_c$ and wavelength $\lambda_c\approx980$~nm tuned to the transition between the $5P_{1/2}$ and $nS_{1/2}$ states.  EIT is observed by measuring the transmission of the 404.8~nm probe beam through the vapor cell as the wavelength of the coupling beam is scanned.

The Doppler effect can have a deleterious impact on measuring EIT in a ladder system.  Whether EIT is observable upon averaging over a thermal velocity distribution depends on the wavelength ordering of the coupling and probe beam \cite{boon1999,urvoy2013}.  If $\lambda_p>\lambda_c$ (the condition satisfied by most previous EIT measurements involving $^{87}$Rb Rydberg states \cite{mohapatra2007,mack2011,tauschinsky2013}), using counter-propagating coupling and probe beams will alleviate the impact of the Doppler effect.  This scheme works because of the behavior of the Doppler shift of the probe beam $\beta f_p$ and the overall EIT Doppler shift $\beta (f_p-f_c)$, where $\beta=v/c$, $v$ is the atomic velocity, and $c$ is the speed of light.  The opposite Doppler shifts for $\lambda_p>\lambda_c$ lead to an observable, albeit broadened, EIT feature. In contrast, if $\lambda_p<\lambda_c$ as in our case, the Doppler effect entirely eliminates the EIT signal.  Hence, we have developed a velocity selection technique for our laser configuration to ensure that the detected EIT signal arises from atoms within a narrow range of velocities, thereby mitigating the impact of Doppler shifts averaged over a thermal velocity distribution.





Our velocity selection scheme involves a strong optical pumping beam that couples the $F=1$ ground state with the $5P_{1/2}$ state.  This resonant pump beam burns a hole near $v=0$ in the $F=1$ velocity distribution [Fig. 1(a)] and transfers atoms to the $5P_{1/2}$ state.  Atoms in the $5P_{1/2}$ state decay to the $F=1$ and $F=2$ ground states through spontaneous emission at a rate $1.07\times10^6$~s$^{-1}$ \cite{sansonetti2008wavelengths}.  Through standard optical pumping, an excess of atoms near $v=0$ will be created in the velocity distribution of the $F=2$ state.  If the intensity of the optical pumping beam is chopped with a period much slower than the the time for inter-atomic collisions to redistribute the population imbalance between hyperfine states, the population hole and peak near $v=0$ [Fig. 1(a)] will be modulated at the chopping frequency.  The signal from the $v=0$ atoms is then reconstructed by measuring the absorption of the probe beam synchronously with the chopping frequency via a lock-in amplifier.

A more general analysis reveals that the demodulated signal arises from two distinct velocity classes.  In the rest frame of an atom with velocity $v=\beta c$, the Doppler-shifted frequency of the probe beam is $\tilde{f}_p=(1-\beta)f_p$ and $\tilde{f}_{op}=(1+\beta)f_{op}$ for the optical pumping beam [Fig. 1(b)].  Since the demodulated signal is only non-zero for changes in the transmission of the probe caused by the optical pumping beam, the atom must be resonant with both beams.  Hence, the demodulated signal is derived from atoms with velocities that satisfy $\left|(1+\beta)f_{op}-(1-\beta)f_p\right|=f_{hp}$, or $\beta=\left[\pm f_{hp}-(f_{op}-f_p)\right]$.  Under the condition $f_{op}-f_{p}=f_{hp}$ (which is enforced in our experiment using an acousto-optic modulator), the velocity classes that give rise to a signal are $\beta=0$ and $\beta=-2f_{hp}/(f_p+f_{op})\approx-f_{hp}/f_p$.  The latter case corresponds to a velocity for which the optical pumping and probe beams in the atomic rest frame are exchanged compared with the zero-velocity case (inset of Fig. 1(a)), and the probe beam is resonant with the $4S,F=1\rightarrow5P_{1/2}$ transition.

EIT for the probe beam is achieved when the overall detuning vanishes in the atomic rest frame, i.e., when $\tilde{f}_c+\tilde{f}_p$ is equal to the frequency difference $f_0$ between the $4S_{1/2}$ and $nS_{1/2}$ states, where $\tilde{f}_c$ is the frequency of the coupling beam in the rest frame of the atom.  This condition for $\beta=0$ is $f_c+f_p=f_0$ and for $\beta=-f_{hp}/f_p$ is $f_p(1-\beta)+f_c(1+\beta)=f_0+f_{hp}$.  With the probe beam stabilized to the $4S_{1/2},F=2\rightarrow5P_{1/2}$ transition, EIT will be observed for $f_c=f_0-f_p$ (for the $v=0$ atoms) and for $f_c=(f_0-f_p)/(1-f_{hp}/f_p)\approx f_0-f_p-190$~MHz (for the atoms with $v= -c f_{hp}/f_p$).  Two EIT features should therefore appear, separated by $190$~MHz.

A schematic of our experimental apparatus is shown in Fig. 2.  The coupling, optical pumping, and probe beam are spatially overlapped and propagate through a potassium vapor cell heated to 90$^\circ$C.  The probe and optical pumping beams (with waists 190~$\mu$m and 120~$\mu$m and powers 70~$\mu$W and 300~$\mu$W, respectively) are derived from the same external cavity diode laser (ECDL); the optics used to frequency lock this laser to the $4S,F=2\rightarrow5P_{1/2}$ transition are not shown.  The probe beam is shifted by $f_{hp}\approx460$~MHz from the optical pumping beam, and its power is chopped at 25~kHz using an acousto-optic modulator.  A separate ECDL is used to generate the 15~mW near-infrared coupling beam, which is weakly focused to a 260~$\mu$m waist in the cell.  The frequency of the coupling beam can be scanned smoothly over approximately 1 GHz using a piezo-electric transducer in the ECDL.  The power of the coupling beam is modulated at 1.27~kHz using a chopper wheel.  The transmitted probe power is measured using a photodetector, and the EIT signal is derived from double-demodulation at 25~kHz and 1.27~kHz via a mixer and lock-in amplifier.  This demodulation scheme detects changes in the transmitted probe power induced by the coupling beam for atomic velocities selected by the optical pumping beam.


\begin{figure}[!htp]
\includegraphics[width=0.8\columnwidth]{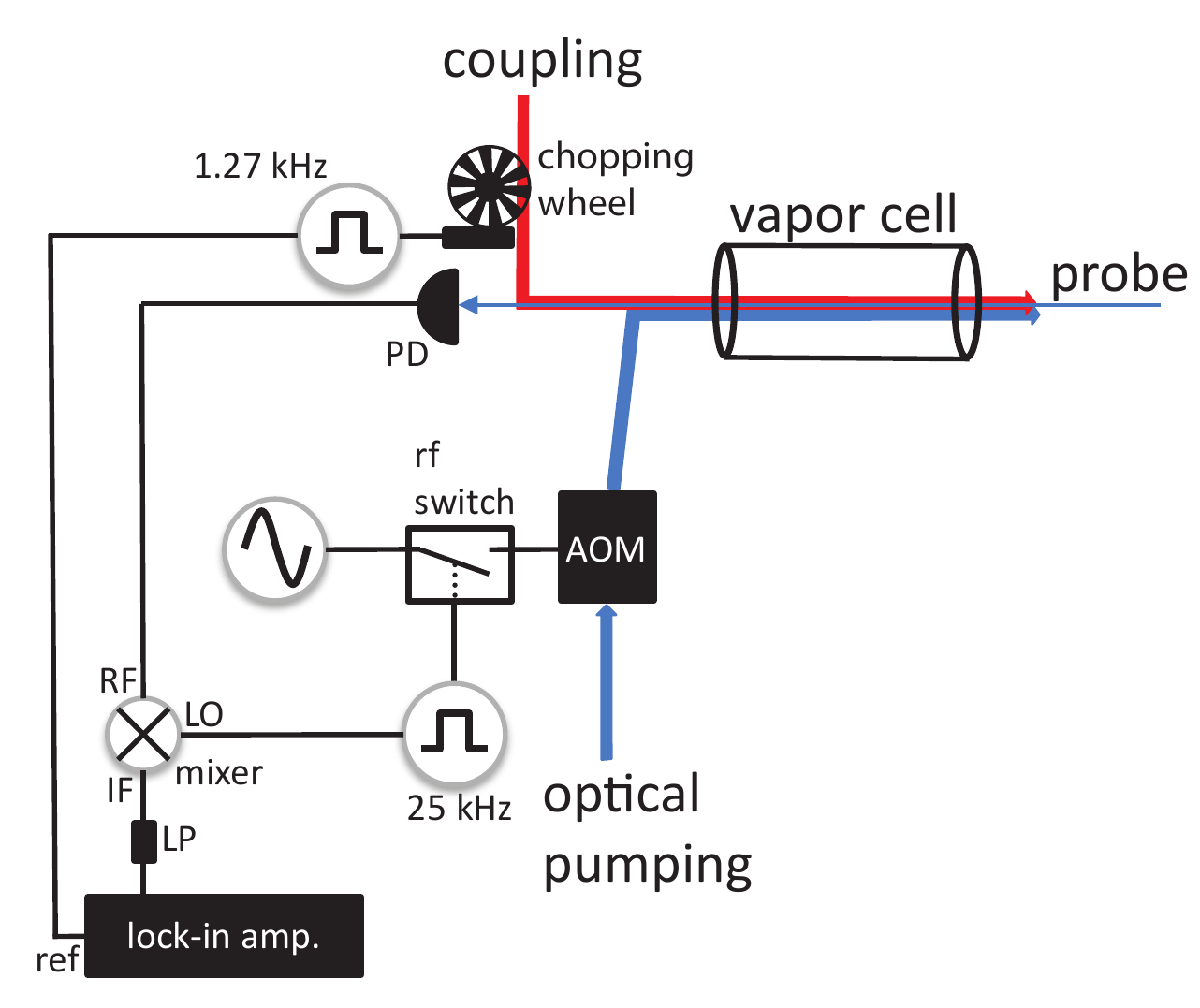}
\caption{Schematic of the EIT measurement apparatus. The coupling beam counter-propagates with the probe beam and co-propagates with the optical pumping beam through a vapor cell. An acousto-optic modulator (AOM) shifts the frequency of optical pumping beam such that the frequency difference with the probe beam matches the ground-state hyperfine splitting. A 25~kHz square-wave generator and radio-frequency (rf) switch is used to modulate the signal generator that drives the AOM and chop the optical pumping beam. A optical chopping wheel modulates the coupling beam at 1.27 kHz. The transmitted power of the probe beam is measured using a photodetector (PD).  The PD signal is doubly demodulated, first using a mixer followed by a low-pass (LP) filter, and then using a lock-in amplifier.}
\label{fig:setup}
\end{figure}



To measure EIT, we tune the near-infrared coupling laser close to the frequency resonant with the $5P_{1/2}\rightarrow nS_{1/2}$ transition predicted by previous measurements of the $4S\rightarrow 5P_{1/2}$ \cite{johansson1972} and $4S\rightarrow nS$ transitions \cite{lorenzen1981}.  The frequency of the coupling laser is scanned and measured using a high-resolution wavemeter (Bristol 621A) while the signal from the lock-in amplifier, which is proportional to changes in the transmitted probe power, is recorded.  Typical data for $n=28$ are shown in Fig. 3. The lock-in amplifier signal is shown vs. the difference in frequency between the coupling laser and the predicted $5P_{1/2}\rightarrow nS$ transition frequency based on Refs. \citenum{lorenzen1981} and \citenum{johansson1972}.  We assume that the wavelength measured in Ref. \citenum{johansson1972} corresponds to the frequency labeled $f_{5P_{1/2}}$ in Fig. 1a.  The two pairs of peaks evident in Fig. 3 derive from the two velocity classes that contribute to the EIT signal, whereas the doublet structure arises from EIT through the $F'=1$ and $F'=2$ hyperfine states in the excited $5P_{1/2}$ electronic state.  The data are fit to a sum of four Gaussian functions, which, in this case, give center frequencies of $(-0.7600 \pm 0.0003)$, $(-0.737 \pm 0.001)$, $(-0.5711 \pm 0.0003)$, and $(-0.5510 \pm 0.0005)$~GHz.  We estimate that drift in the laser lock for the 404.8~nm laser adds a 1~MHz uncertainty to these frequencies; Zeeman and AC Stark shifts are negligible at this level.  The $20\pm2$~MHz difference between the closely spaced peaks is consistent with the $18.1\pm0.2$~MHz $5P_{1/2}$ hyperfine splitting \cite{arimondo1977}. The approximately 190~MHz difference between the pairs is consistent with the shift in the resonant EIT coupling frequency between atoms with $v=0$ and $v=-c f_{hp}/f_p$.

\begin{figure}[!htp]
\includegraphics[width=0.8\columnwidth]{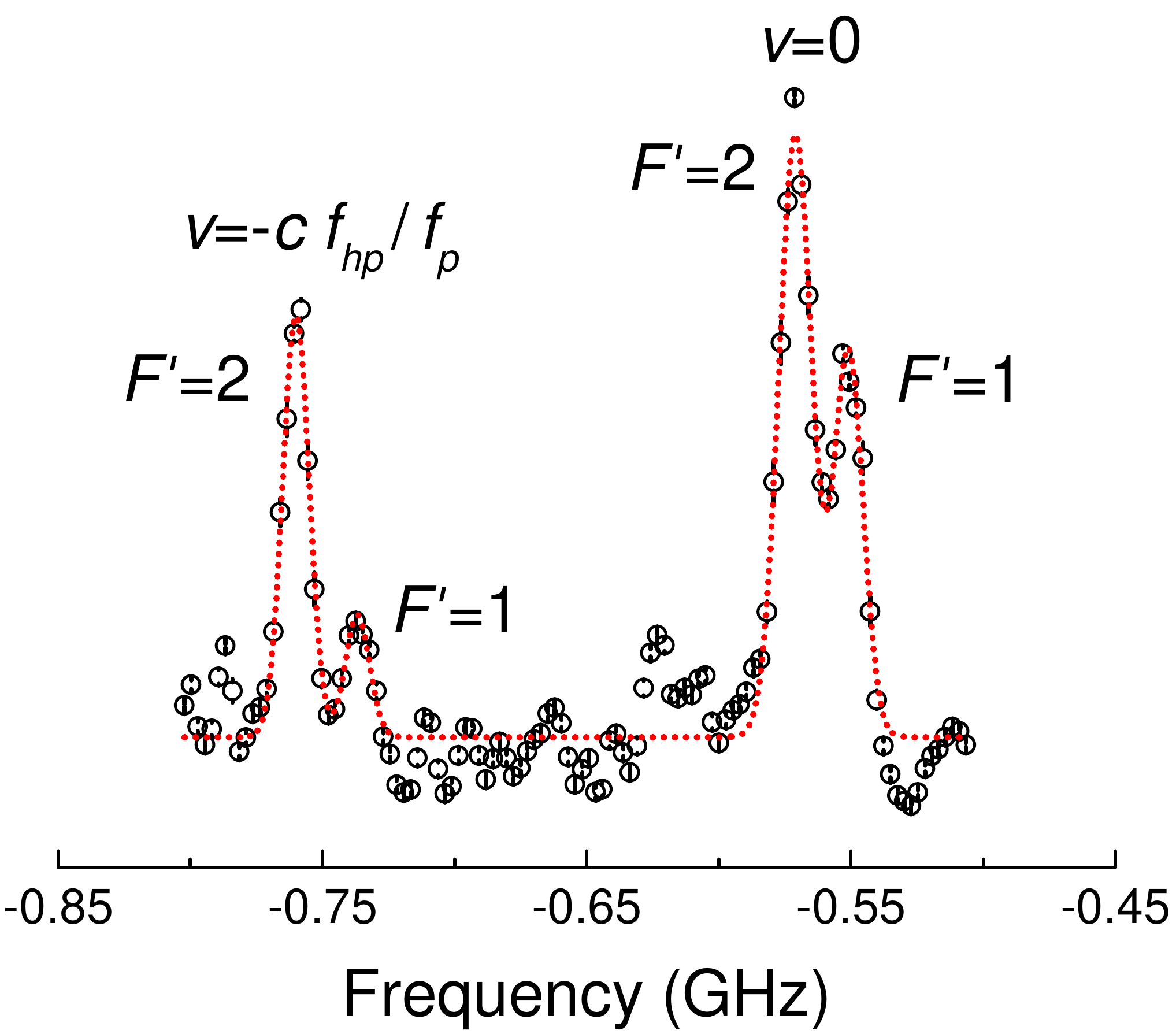}
\caption{EIT spectroscopy of the $28S$ state.  The signal from the lock-in amplifier, which is proportional to changes in the probe transmission, is shown as the frequency of the coupling laser is scanned. The abscissa is the frequency difference of the coupling laser relative to the $5P_{1/2}\rightarrow nS$ transition predicted by previous measurements of the $4S\rightarrow 5P_{1/2}$ and $4S\rightarrow nS$ transitions \cite{lorenzen1981,johansson1972}.  Each point is an average of 5 measurements, and the error bars are the standard error of the mean.  The dotted line is a fit to the sum of four Gaussian functions.}
\label{fig:Spectro}
\end{figure}



\begin{figure}[!htp]
\includegraphics[width=0.8\columnwidth]{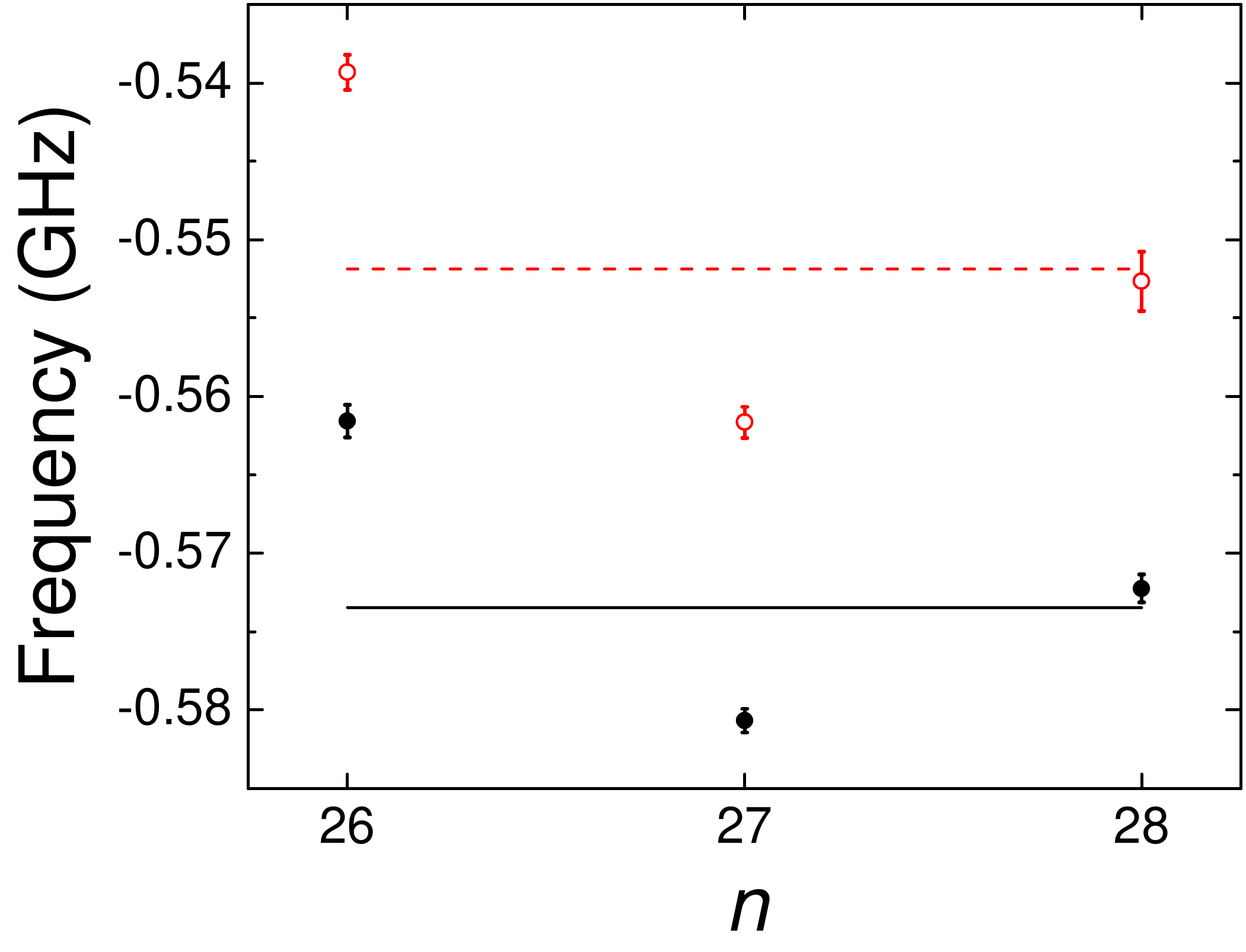}
\caption{The measured frequency deviation for the $5P_{1/2} \rightarrow nS_{1/2}$ transition with principal quantum numbers $n=26$, 27 and 28. The solid black (hollow red) points are for the transition to the $5P,\;F'=1$ ($5P,\;F'=2$) state. The black solid and red dashed lines are the weighted average of each set of three points for the transitions to the $5P,\;F'=1$ and $5P,\;F'=2$ states, respectively.  The error bars show the uncertainty from the fits to data such as those in Fig. 3.  The 60~MHz uncertainty from the accuracy of the wavemeter is not included in the error bars.}
\label{fig:Freq}
\end{figure}

We measure these EIT features for $n=26$, 27, and 28, which spans 0.86~THz in the coupling laser frequency.  Fig.~\ref{fig:Freq} shows the center frequencies of the EIT peaks (as deviations from the frequency predicted by Refs. \citenum{lorenzen1981} and \citenum{johansson1972}) for $v=0$ obtained from Gaussian fits such as those shown in Fig.~3.  The weighted average for the frequency deviation of the transition to the $5P,\;F'=1$ and $5P,\;F'=2$ states is $-573 \pm 0.005$ GHz and $-552 \pm 0.007$ GHz, respectively.  The individual measurements are consistent with these average values within the specified 10 MHz repeatability of the wavemeter.  The most significant contribution to the overall measurement uncertainty is the 60~MHz accuracy of the wavemeter used to measure the absolute frequency of the coupling laser.  We have verified that the wavemeter is accurate at this level for the $^{87}$Rb and $^{40}$K D2 transitions.  We assign an additional 20~MHz uncertainty to account for the unresolved hyperfine structure of the $5P_{1/2}$ state in the setup used to stabilize the wavelength of the 404.8~nm laser, making the overall uncertainty 80~MHz.


Based on our measurement, the $4S_{1/2}\rightarrow5P_{1/2}$ transition frequency (as labeled $f_{5P_{1/2}}$ in Fig. 1a) should be shifted by $560\pm80$~MHz from the currently accepted value \cite{johansson1972,nist}, giving $740529.36\pm0.08$~GHz for the absolute frequency of this transition.   Our measurement represents a factor of two improvement in the present $\pm150$~MHz reported uncertainty for this transition \cite{johansson1972}.  Our results are not consistent with the currently accepted value of this transition, which was determined using a grating-based spectrometer and hollow-cathode lamp \cite{johansson1972}, and was therefore not a measurement of the absolute frequency.   Also, the ground-state hyperfine structure was not accounted for in Ref. \citenum{johansson1972}, potentially introducing large systematic errors.  Furthermore, the source of the $\pm150$~MHz reported uncertainty and a justification of that value were not explained in Ref.~\cite{johansson1972}.

In conclusion, we have developed a simple method to mitigate the impact of the Doppler effect on EIT spectroscopy in a vapor cell for high--low probe--coupling wavelength ordering.   We have determined an improved value for the frequency of the $4S_{1/2}\rightarrow5P_{1/2}$ transition in $^{39}$K, which is useful for Rydberg dressing and may have an impact on understanding aeronomic nightglow \cite{slanger2000accurate} and astronomical studies of cool or dust-obscured objects \cite{civivs2012potassium}.  Our  method can be adapted to frequency stabilize the near-infrared coupling laser to the $5P_{1/2}\rightarrow nS_{1/2}$ transition via frequency modulation spectroscopy \cite{abel2009}. Future work includes applying 404.8~nm and approximately 980~nm light to $^{40}$K atoms trapped in an optical lattice to investigate Rydberg dressing.

\begin{acknowledgments}
The authors acknowledge funding from the National Science Foundation (grant PHY15-05468) and the Army Research Office (grant W911NF-12-1-0462).  W. Xu acknowledges assistance with building the apparatus from W. R. McGehee.
\end{acknowledgments}

\clearpage

\bibliography{RefRydberg}

\begin{thebibliography}{36}%
\makeatletter
\providecommand \@ifxundefined [1]{%
 \@ifx{#1\undefined}
}%
\providecommand \@ifnum [1]{%
 \ifnum #1\expandafter \@firstoftwo
 \else \expandafter \@secondoftwo
 \fi
}%
\providecommand \@ifx [1]{%
 \ifx #1\expandafter \@firstoftwo
 \else \expandafter \@secondoftwo
 \fi
}%
\providecommand \natexlab [1]{#1}%
\providecommand \enquote  [1]{``#1''}%
\providecommand \bibnamefont  [1]{#1}%
\providecommand \bibfnamefont [1]{#1}%
\providecommand \citenamefont [1]{#1}%
\providecommand \href@noop [0]{\@secondoftwo}%
\providecommand \href [0]{\begingroup \@sanitize@url \@href}%
\providecommand \@href[1]{\@@startlink{#1}\@@href}%
\providecommand \@@href[1]{\endgroup#1\@@endlink}%
\providecommand \@sanitize@url [0]{\catcode `\\12\catcode `\$12\catcode
  `\&12\catcode `\#12\catcode `\^12\catcode `\_12\catcode `\%12\relax}%
\providecommand \@@startlink[1]{}%
\providecommand \@@endlink[0]{}%
\providecommand \url  [0]{\begingroup\@sanitize@url \@url }%
\providecommand \@url [1]{\endgroup\@href {#1}{\urlprefix }}%
\providecommand \urlprefix  [0]{URL }%
\providecommand \Eprint [0]{\href }%
\providecommand \doibase [0]{http://dx.doi.org/}%
\providecommand \selectlanguage [0]{\@gobble}%
\providecommand \bibinfo  [0]{\@secondoftwo}%
\providecommand \bibfield  [0]{\@secondoftwo}%
\providecommand \translation [1]{[#1]}%
\providecommand \BibitemOpen [0]{}%
\providecommand \bibitemStop [0]{}%
\providecommand \bibitemNoStop [0]{.\EOS\space}%
\providecommand \EOS [0]{\spacefactor3000\relax}%
\providecommand \BibitemShut  [1]{\csname bibitem#1\endcsname}%
\let\auto@bib@innerbib\@empty
\bibitem [{\citenamefont {Gorshkov}\ \emph {et~al.}(2011)\citenamefont
  {Gorshkov}, \citenamefont {Manmana}, \citenamefont {Chen}, \citenamefont
  {Ye}, \citenamefont {Demler}, \citenamefont {Lukin},\ and\ \citenamefont
  {Rey}}]{Gorshkov2011}%
  \BibitemOpen
  \bibfield  {author} {\bibinfo {author} {\bibfnamefont {A.~V.}\ \bibnamefont
  {Gorshkov}}, \bibinfo {author} {\bibfnamefont {S.~R.}\ \bibnamefont
  {Manmana}}, \bibinfo {author} {\bibfnamefont {G.}~\bibnamefont {Chen}},
  \bibinfo {author} {\bibfnamefont {J.}~\bibnamefont {Ye}}, \bibinfo {author}
  {\bibfnamefont {E.}~\bibnamefont {Demler}}, \bibinfo {author} {\bibfnamefont
  {M.~D.}\ \bibnamefont {Lukin}}, \ and\ \bibinfo {author} {\bibfnamefont
  {A.~M.}\ \bibnamefont {Rey}},\ }\href {\doibase
  10.1103/PhysRevLett.107.115301} {\bibfield  {journal} {\bibinfo  {journal}
  {Phys. Rev. Lett.}\ }\textbf {\bibinfo {volume} {107}},\ \bibinfo {pages}
  {115301} (\bibinfo {year} {2011})}\BibitemShut {NoStop}%
\bibitem [{\citenamefont {Moses}\ \emph {et~al.}(2015)\citenamefont {Moses},
  \citenamefont {Covey}, \citenamefont {Miecnikowski}, \citenamefont {Yan},
  \citenamefont {Gadway}, \citenamefont {Ye},\ and\ \citenamefont
  {Jin}}]{moses2015}%
  \BibitemOpen
  \bibfield  {author} {\bibinfo {author} {\bibfnamefont {S.~A.}\ \bibnamefont
  {Moses}}, \bibinfo {author} {\bibfnamefont {J.~P.}\ \bibnamefont {Covey}},
  \bibinfo {author} {\bibfnamefont {M.~T.}\ \bibnamefont {Miecnikowski}},
  \bibinfo {author} {\bibfnamefont {B.}~\bibnamefont {Yan}}, \bibinfo {author}
  {\bibfnamefont {B.}~\bibnamefont {Gadway}}, \bibinfo {author} {\bibfnamefont
  {J.}~\bibnamefont {Ye}}, \ and\ \bibinfo {author} {\bibfnamefont {D.~S.}\
  \bibnamefont {Jin}},\ }\href@noop {} {\bibfield  {journal} {\bibinfo
  {journal} {arXiv:1507.02377}\ } (\bibinfo {year} {2015})}\BibitemShut
  {NoStop}%
\bibitem [{\citenamefont {Cooper}\ and\ \citenamefont
  {Shlyapnikov}(2009)}]{Cooper2009}%
  \BibitemOpen
  \bibfield  {author} {\bibinfo {author} {\bibfnamefont {N.~R.}\ \bibnamefont
  {Cooper}}\ and\ \bibinfo {author} {\bibfnamefont {G.~V.}\ \bibnamefont
  {Shlyapnikov}},\ }\href {\doibase 10.1103/PhysRevLett.103.155302} {\bibfield
  {journal} {\bibinfo  {journal} {Phys. Rev. Lett.}\ }\textbf {\bibinfo
  {volume} {103}},\ \bibinfo {pages} {155302} (\bibinfo {year}
  {2009})}\BibitemShut {NoStop}%
\bibitem [{\citenamefont {Yan}\ \emph {et~al.}(2013)\citenamefont {Yan},
  \citenamefont {Moses}, \citenamefont {Gadway}, \citenamefont {Covey},
  \citenamefont {Hazzard}, \citenamefont {Rey}, \citenamefont {Jin},\ and\
  \citenamefont {Ye}}]{Yan2013}%
  \BibitemOpen
  \bibfield  {author} {\bibinfo {author} {\bibfnamefont {B.}~\bibnamefont
  {Yan}}, \bibinfo {author} {\bibfnamefont {S.~A.}\ \bibnamefont {Moses}},
  \bibinfo {author} {\bibfnamefont {B.}~\bibnamefont {Gadway}}, \bibinfo
  {author} {\bibfnamefont {J.~P.}\ \bibnamefont {Covey}}, \bibinfo {author}
  {\bibfnamefont {K.~R.}\ \bibnamefont {Hazzard}}, \bibinfo {author}
  {\bibfnamefont {A.~M.}\ \bibnamefont {Rey}}, \bibinfo {author} {\bibfnamefont
  {D.~S.}\ \bibnamefont {Jin}}, \ and\ \bibinfo {author} {\bibfnamefont
  {J.}~\bibnamefont {Ye}},\ }\href@noop {} {\bibfield  {journal} {\bibinfo
  {journal} {Nature}\ }\textbf {\bibinfo {volume} {501}},\ \bibinfo {pages}
  {521} (\bibinfo {year} {2013})}\BibitemShut {NoStop}%
\bibitem [{\citenamefont {Lu}\ \emph {et~al.}(2012)\citenamefont {Lu},
  \citenamefont {Burdick},\ and\ \citenamefont {Lev}}]{Lu2012}%
  \BibitemOpen
  \bibfield  {author} {\bibinfo {author} {\bibfnamefont {M.}~\bibnamefont
  {Lu}}, \bibinfo {author} {\bibfnamefont {N.~Q.}\ \bibnamefont {Burdick}}, \
  and\ \bibinfo {author} {\bibfnamefont {B.~L.}\ \bibnamefont {Lev}},\ }\href
  {\doibase 10.1103/PhysRevLett.108.215301} {\bibfield  {journal} {\bibinfo
  {journal} {Phys. Rev. Lett.}\ }\textbf {\bibinfo {volume} {108}},\ \bibinfo
  {pages} {215301} (\bibinfo {year} {2012})}\BibitemShut {NoStop}%
\bibitem [{\citenamefont {Stuhler}\ \emph {et~al.}(2005)\citenamefont
  {Stuhler}, \citenamefont {Griesmaier}, \citenamefont {Koch}, \citenamefont
  {Fattori}, \citenamefont {Pfau}, \citenamefont {Giovanazzi}, \citenamefont
  {Pedri},\ and\ \citenamefont {Santos}}]{stuhler2005}%
  \BibitemOpen
  \bibfield  {author} {\bibinfo {author} {\bibfnamefont {J.}~\bibnamefont
  {Stuhler}}, \bibinfo {author} {\bibfnamefont {A.}~\bibnamefont {Griesmaier}},
  \bibinfo {author} {\bibfnamefont {T.}~\bibnamefont {Koch}}, \bibinfo {author}
  {\bibfnamefont {M.}~\bibnamefont {Fattori}}, \bibinfo {author} {\bibfnamefont
  {T.}~\bibnamefont {Pfau}}, \bibinfo {author} {\bibfnamefont {S.}~\bibnamefont
  {Giovanazzi}}, \bibinfo {author} {\bibfnamefont {P.}~\bibnamefont {Pedri}}, \
  and\ \bibinfo {author} {\bibfnamefont {L.}~\bibnamefont {Santos}},\
  }\href@noop {} {\bibfield  {journal} {\bibinfo  {journal} {Phys. Rev. Lett.}\
  }\textbf {\bibinfo {volume} {95}},\ \bibinfo {pages} {150406} (\bibinfo
  {year} {2005})}\BibitemShut {NoStop}%
\bibitem [{\citenamefont {Aikawa}\ \emph {et~al.}(2012)\citenamefont {Aikawa},
  \citenamefont {Frisch}, \citenamefont {Mark}, \citenamefont {Baier},
  \citenamefont {Rietzler}, \citenamefont {Grimm},\ and\ \citenamefont
  {Ferlaino}}]{aikawa2012}%
  \BibitemOpen
  \bibfield  {author} {\bibinfo {author} {\bibfnamefont {K.}~\bibnamefont
  {Aikawa}}, \bibinfo {author} {\bibfnamefont {A.}~\bibnamefont {Frisch}},
  \bibinfo {author} {\bibfnamefont {M.}~\bibnamefont {Mark}}, \bibinfo {author}
  {\bibfnamefont {S.}~\bibnamefont {Baier}}, \bibinfo {author} {\bibfnamefont
  {A.}~\bibnamefont {Rietzler}}, \bibinfo {author} {\bibfnamefont
  {R.}~\bibnamefont {Grimm}}, \ and\ \bibinfo {author} {\bibfnamefont
  {F.}~\bibnamefont {Ferlaino}},\ }\href@noop {} {\bibfield  {journal}
  {\bibinfo  {journal} {Phys. Rev. Lett.}\ }\textbf {\bibinfo {volume} {108}},\
  \bibinfo {pages} {210401} (\bibinfo {year} {2012})}\BibitemShut {NoStop}%
\bibitem [{\citenamefont {Viteau}\ \emph {et~al.}(2011)\citenamefont {Viteau},
  \citenamefont {Bason}, \citenamefont {Radogostowicz}, \citenamefont
  {Malossi}, \citenamefont {Ciampini}, \citenamefont {Morsch},\ and\
  \citenamefont {Arimondo}}]{Viteau2011}%
  \BibitemOpen
  \bibfield  {author} {\bibinfo {author} {\bibfnamefont {M.}~\bibnamefont
  {Viteau}}, \bibinfo {author} {\bibfnamefont {M.~G.}\ \bibnamefont {Bason}},
  \bibinfo {author} {\bibfnamefont {J.}~\bibnamefont {Radogostowicz}}, \bibinfo
  {author} {\bibfnamefont {N.}~\bibnamefont {Malossi}}, \bibinfo {author}
  {\bibfnamefont {D.}~\bibnamefont {Ciampini}}, \bibinfo {author}
  {\bibfnamefont {O.}~\bibnamefont {Morsch}}, \ and\ \bibinfo {author}
  {\bibfnamefont {E.}~\bibnamefont {Arimondo}},\ }\href {\doibase
  10.1103/PhysRevLett.107.060402} {\bibfield  {journal} {\bibinfo  {journal}
  {Phys. Rev. Lett.}\ }\textbf {\bibinfo {volume} {107}},\ \bibinfo {pages}
  {060402} (\bibinfo {year} {2011})}\BibitemShut {NoStop}%
\bibitem [{\citenamefont {Zeiher}\ \emph {et~al.}(2015)\citenamefont {Zeiher},
  \citenamefont {Schau{\ss}}, \citenamefont {Hild}, \citenamefont {Macr{\`\i}},
  \citenamefont {Bloch},\ and\ \citenamefont {Gross}}]{zeiher2015}%
  \BibitemOpen
  \bibfield  {author} {\bibinfo {author} {\bibfnamefont {J.}~\bibnamefont
  {Zeiher}}, \bibinfo {author} {\bibfnamefont {P.}~\bibnamefont {Schau{\ss}}},
  \bibinfo {author} {\bibfnamefont {S.}~\bibnamefont {Hild}}, \bibinfo {author}
  {\bibfnamefont {T.}~\bibnamefont {Macr{\`\i}}}, \bibinfo {author}
  {\bibfnamefont {I.}~\bibnamefont {Bloch}}, \ and\ \bibinfo {author}
  {\bibfnamefont {C.}~\bibnamefont {Gross}},\ }\href@noop {} {\bibfield
  {journal} {\bibinfo  {journal} {Phys. Rev. X}\ }\textbf {\bibinfo {volume}
  {5}},\ \bibinfo {pages} {031015} (\bibinfo {year} {2015})}\BibitemShut
  {NoStop}%
\bibitem [{\citenamefont {Schau{\ss}}\ \emph {et~al.}(2012)\citenamefont
  {Schau{\ss}}, \citenamefont {Cheneau}, \citenamefont {Endres}, \citenamefont
  {Fukuhara}, \citenamefont {Hild}, \citenamefont {Omran}, \citenamefont
  {Pohl}, \citenamefont {Gross}, \citenamefont {Kuhr},\ and\ \citenamefont
  {Bloch}}]{Schau2012}%
  \BibitemOpen
  \bibfield  {author} {\bibinfo {author} {\bibfnamefont {P.}~\bibnamefont
  {Schau{\ss}}}, \bibinfo {author} {\bibfnamefont {M.}~\bibnamefont {Cheneau}},
  \bibinfo {author} {\bibfnamefont {M.}~\bibnamefont {Endres}}, \bibinfo
  {author} {\bibfnamefont {T.}~\bibnamefont {Fukuhara}}, \bibinfo {author}
  {\bibfnamefont {S.}~\bibnamefont {Hild}}, \bibinfo {author} {\bibfnamefont
  {A.}~\bibnamefont {Omran}}, \bibinfo {author} {\bibfnamefont
  {T.}~\bibnamefont {Pohl}}, \bibinfo {author} {\bibfnamefont {C.}~\bibnamefont
  {Gross}}, \bibinfo {author} {\bibfnamefont {S.}~\bibnamefont {Kuhr}}, \ and\
  \bibinfo {author} {\bibfnamefont {I.}~\bibnamefont {Bloch}},\ }\href@noop {}
  {\bibfield  {journal} {\bibinfo  {journal} {Nature}\ }\textbf {\bibinfo
  {volume} {491}},\ \bibinfo {pages} {87} (\bibinfo {year} {2012})}\BibitemShut
  {NoStop}%
\bibitem [{\citenamefont {Schau{\ss}}\ \emph {et~al.}(2015)\citenamefont
  {Schau{\ss}}, \citenamefont {Zeiher}, \citenamefont {Fukuhara}, \citenamefont
  {Hild}, \citenamefont {Cheneau}, \citenamefont {Macr{\`\i}}, \citenamefont
  {Pohl}, \citenamefont {Bloch},\ and\ \citenamefont {Gro{\ss}}}]{schauss2015}%
  \BibitemOpen
  \bibfield  {author} {\bibinfo {author} {\bibfnamefont {P.}~\bibnamefont
  {Schau{\ss}}}, \bibinfo {author} {\bibfnamefont {J.}~\bibnamefont {Zeiher}},
  \bibinfo {author} {\bibfnamefont {T.}~\bibnamefont {Fukuhara}}, \bibinfo
  {author} {\bibfnamefont {S.}~\bibnamefont {Hild}}, \bibinfo {author}
  {\bibfnamefont {M.}~\bibnamefont {Cheneau}}, \bibinfo {author} {\bibfnamefont
  {T.}~\bibnamefont {Macr{\`\i}}}, \bibinfo {author} {\bibfnamefont
  {T.}~\bibnamefont {Pohl}}, \bibinfo {author} {\bibfnamefont {I.}~\bibnamefont
  {Bloch}}, \ and\ \bibinfo {author} {\bibfnamefont {C.}~\bibnamefont
  {Gro{\ss}}},\ }\href@noop {} {\bibfield  {journal} {\bibinfo  {journal}
  {Science}\ }\textbf {\bibinfo {volume} {347}},\ \bibinfo {pages} {1455}
  (\bibinfo {year} {2015})}\BibitemShut {NoStop}%
\bibitem [{\citenamefont {Macr\`\i}\ and\ \citenamefont
  {Pohl}(2014)}]{Macr2014}%
  \BibitemOpen
  \bibfield  {author} {\bibinfo {author} {\bibfnamefont {T.}~\bibnamefont
  {Macr\`\i}}\ and\ \bibinfo {author} {\bibfnamefont {T.}~\bibnamefont
  {Pohl}},\ }\href {\doibase 10.1103/PhysRevA.89.011402} {\bibfield  {journal}
  {\bibinfo  {journal} {Phys. Rev. A}\ }\textbf {\bibinfo {volume} {89}},\
  \bibinfo {pages} {011402} (\bibinfo {year} {2014})}\BibitemShut {NoStop}%
\bibitem [{\citenamefont {Johnson}\ and\ \citenamefont
  {Rolston}(2010)}]{johnson2010}%
  \BibitemOpen
  \bibfield  {author} {\bibinfo {author} {\bibfnamefont {J.}~\bibnamefont
  {Johnson}}\ and\ \bibinfo {author} {\bibfnamefont {S.}~\bibnamefont
  {Rolston}},\ }\href@noop {} {\bibfield  {journal} {\bibinfo  {journal} {Phys.
  Rev. A}\ }\textbf {\bibinfo {volume} {82}},\ \bibinfo {pages} {033412}
  (\bibinfo {year} {2010})}\BibitemShut {NoStop}%
\bibitem [{\citenamefont {Saffman}\ \emph {et~al.}(2010)\citenamefont
  {Saffman}, \citenamefont {Walker},\ and\ \citenamefont
  {M{\o}lmer}}]{saffman2010}%
  \BibitemOpen
  \bibfield  {author} {\bibinfo {author} {\bibfnamefont {M.}~\bibnamefont
  {Saffman}}, \bibinfo {author} {\bibfnamefont {T.}~\bibnamefont {Walker}}, \
  and\ \bibinfo {author} {\bibfnamefont {K.}~\bibnamefont {M{\o}lmer}},\
  }\href@noop {} {\bibfield  {journal} {\bibinfo  {journal} {Rev. Mod. Phys.}\
  }\textbf {\bibinfo {volume} {82}},\ \bibinfo {pages} {2313} (\bibinfo {year}
  {2010})}\BibitemShut {NoStop}%
\bibitem [{\citenamefont {Balewski}\ \emph {et~al.}(2014)\citenamefont
  {Balewski}, \citenamefont {Krupp}, \citenamefont {Gaj}, \citenamefont
  {Hofferberth}, \citenamefont {L\"{o}w},\ and\ \citenamefont
  {Pfau}}]{1367-2630-16-6-063012}%
  \BibitemOpen
  \bibfield  {author} {\bibinfo {author} {\bibfnamefont {J.~B.}\ \bibnamefont
  {Balewski}}, \bibinfo {author} {\bibfnamefont {A.~T.}\ \bibnamefont {Krupp}},
  \bibinfo {author} {\bibfnamefont {A.}~\bibnamefont {Gaj}}, \bibinfo {author}
  {\bibfnamefont {S.}~\bibnamefont {Hofferberth}}, \bibinfo {author}
  {\bibfnamefont {R.}~\bibnamefont {L\"{o}w}}, \ and\ \bibinfo {author}
  {\bibfnamefont {T.}~\bibnamefont {Pfau}},\ }\href
  {http://stacks.iop.org/1367-2630/16/i=6/a=063012} {\bibfield  {journal}
  {\bibinfo  {journal} {N. J. Phys.}\ }\textbf {\bibinfo {volume} {16}},\
  \bibinfo {pages} {063012} (\bibinfo {year} {2014})}\BibitemShut {NoStop}%
\bibitem [{\citenamefont {Mattioli}\ \emph {et~al.}(2013)\citenamefont
  {Mattioli}, \citenamefont {Dalmonte}, \citenamefont {Lechner},\ and\
  \citenamefont {Pupillo}}]{mattioli2013}%
  \BibitemOpen
  \bibfield  {author} {\bibinfo {author} {\bibfnamefont {M.}~\bibnamefont
  {Mattioli}}, \bibinfo {author} {\bibfnamefont {M.}~\bibnamefont {Dalmonte}},
  \bibinfo {author} {\bibfnamefont {W.}~\bibnamefont {Lechner}}, \ and\
  \bibinfo {author} {\bibfnamefont {G.}~\bibnamefont {Pupillo}},\ }\href@noop
  {} {\bibfield  {journal} {\bibinfo  {journal} {Phys. Rev. Lett.}\ }\textbf
  {\bibinfo {volume} {111}},\ \bibinfo {pages} {165302} (\bibinfo {year}
  {2013})}\BibitemShut {NoStop}%
\bibitem [{\citenamefont {Li}\ and\ \citenamefont {Sarma}(2015)}]{li2015}%
  \BibitemOpen
  \bibfield  {author} {\bibinfo {author} {\bibfnamefont {X.}~\bibnamefont
  {Li}}\ and\ \bibinfo {author} {\bibfnamefont {S.~D.}\ \bibnamefont {Sarma}},\
  }\href@noop {} {\bibfield  {journal} {\bibinfo  {journal} {Nat. Comm.}\
  }\textbf {\bibinfo {volume} {6}},\ \bibinfo {pages} {7137} (\bibinfo {year}
  {2015})}\BibitemShut {NoStop}%
\bibitem [{\citenamefont {J{\"o}rdens}\ \emph {et~al.}(2008)\citenamefont
  {J{\"o}rdens}, \citenamefont {Strohmaier}, \citenamefont {G{\"u}nter},
  \citenamefont {Moritz},\ and\ \citenamefont {Esslinger}}]{jordens2008mott}%
  \BibitemOpen
  \bibfield  {author} {\bibinfo {author} {\bibfnamefont {R.}~\bibnamefont
  {J{\"o}rdens}}, \bibinfo {author} {\bibfnamefont {N.}~\bibnamefont
  {Strohmaier}}, \bibinfo {author} {\bibfnamefont {K.}~\bibnamefont
  {G{\"u}nter}}, \bibinfo {author} {\bibfnamefont {H.}~\bibnamefont {Moritz}},
  \ and\ \bibinfo {author} {\bibfnamefont {T.}~\bibnamefont {Esslinger}},\
  }\href@noop {} {\bibfield  {journal} {\bibinfo  {journal} {Nature}\ }\textbf
  {\bibinfo {volume} {455}},\ \bibinfo {pages} {204} (\bibinfo {year}
  {2008})}\BibitemShut {NoStop}%
\bibitem [{\citenamefont {Schneider}\ \emph {et~al.}(2008)\citenamefont
  {Schneider}, \citenamefont {Hackerm{\"u}ller}, \citenamefont {Will},
  \citenamefont {Best}, \citenamefont {Bloch}, \citenamefont {Costi},
  \citenamefont {Helmes}, \citenamefont {Rasch},\ and\ \citenamefont
  {Rosch}}]{schneider2008metallic}%
  \BibitemOpen
  \bibfield  {author} {\bibinfo {author} {\bibfnamefont {U.}~\bibnamefont
  {Schneider}}, \bibinfo {author} {\bibfnamefont {L.}~\bibnamefont
  {Hackerm{\"u}ller}}, \bibinfo {author} {\bibfnamefont {S.}~\bibnamefont
  {Will}}, \bibinfo {author} {\bibfnamefont {T.}~\bibnamefont {Best}}, \bibinfo
  {author} {\bibfnamefont {I.}~\bibnamefont {Bloch}}, \bibinfo {author}
  {\bibfnamefont {T.}~\bibnamefont {Costi}}, \bibinfo {author} {\bibfnamefont
  {R.}~\bibnamefont {Helmes}}, \bibinfo {author} {\bibfnamefont
  {D.}~\bibnamefont {Rasch}}, \ and\ \bibinfo {author} {\bibfnamefont
  {A.}~\bibnamefont {Rosch}},\ }\href@noop {} {\bibfield  {journal} {\bibinfo
  {journal} {Science}\ }\textbf {\bibinfo {volume} {322}},\ \bibinfo {pages}
  {1520} (\bibinfo {year} {2008})}\BibitemShut {NoStop}%
\bibitem [{\citenamefont {Lorenzen}\ \emph {et~al.}(1981)\citenamefont
  {Lorenzen}, \citenamefont {Niemax},\ and\ \citenamefont
  {Pendrill}}]{lorenzen1981}%
  \BibitemOpen
  \bibfield  {author} {\bibinfo {author} {\bibfnamefont {C.-J.}\ \bibnamefont
  {Lorenzen}}, \bibinfo {author} {\bibfnamefont {K.}~\bibnamefont {Niemax}}, \
  and\ \bibinfo {author} {\bibfnamefont {L.}~\bibnamefont {Pendrill}},\
  }\href@noop {} {\bibfield  {journal} {\bibinfo  {journal} {Opt. Comm.}\
  }\textbf {\bibinfo {volume} {39}},\ \bibinfo {pages} {370} (\bibinfo {year}
  {1981})}\BibitemShut {NoStop}%
\bibitem [{\citenamefont {Johansson}\ and\ \citenamefont
  {Svendenius}(1972)}]{johansson1972}%
  \BibitemOpen
  \bibfield  {author} {\bibinfo {author} {\bibfnamefont {I.}~\bibnamefont
  {Johansson}}\ and\ \bibinfo {author} {\bibfnamefont {N.}~\bibnamefont
  {Svendenius}},\ }\href@noop {} {\bibfield  {journal} {\bibinfo  {journal}
  {Phys. Scripta}\ }\textbf {\bibinfo {volume} {5}},\ \bibinfo {pages} {129}
  (\bibinfo {year} {1972})}\BibitemShut {NoStop}%
\bibitem [{\citenamefont {Fleischhauer}\ \emph {et~al.}(2005)\citenamefont
  {Fleischhauer}, \citenamefont {Imamoglu},\ and\ \citenamefont
  {Marangos}}]{Fleischhauer2005}%
  \BibitemOpen
  \bibfield  {author} {\bibinfo {author} {\bibfnamefont {M.}~\bibnamefont
  {Fleischhauer}}, \bibinfo {author} {\bibfnamefont {A.}~\bibnamefont
  {Imamoglu}}, \ and\ \bibinfo {author} {\bibfnamefont {J.~P.}\ \bibnamefont
  {Marangos}},\ }\href {\doibase 10.1103/RevModPhys.77.633} {\bibfield
  {journal} {\bibinfo  {journal} {Rev. Mod. Phys.}\ }\textbf {\bibinfo {volume}
  {77}},\ \bibinfo {pages} {633} (\bibinfo {year} {2005})}\BibitemShut
  {NoStop}%
\bibitem [{\citenamefont {Behrle}\ \emph {et~al.}(2011)\citenamefont {Behrle},
  \citenamefont {Koschorreck},\ and\ \citenamefont {K{\"o}hl}}]{behrle2011}%
  \BibitemOpen
  \bibfield  {author} {\bibinfo {author} {\bibfnamefont {A.}~\bibnamefont
  {Behrle}}, \bibinfo {author} {\bibfnamefont {M.}~\bibnamefont {Koschorreck}},
  \ and\ \bibinfo {author} {\bibfnamefont {M.}~\bibnamefont {K{\"o}hl}},\
  }\href@noop {} {\bibfield  {journal} {\bibinfo  {journal} {Phys. Rev. A}\
  }\textbf {\bibinfo {volume} {83}},\ \bibinfo {pages} {052507} (\bibinfo
  {year} {2011})}\BibitemShut {NoStop}%
\bibitem [{\citenamefont {Niemax}\ and\ \citenamefont
  {Pendrill}(1980)}]{niemax1980}%
  \BibitemOpen
  \bibfield  {author} {\bibinfo {author} {\bibfnamefont {K.}~\bibnamefont
  {Niemax}}\ and\ \bibinfo {author} {\bibfnamefont {L.}~\bibnamefont
  {Pendrill}},\ }\href@noop {} {\bibfield  {journal} {\bibinfo  {journal} {J.
  Phys. B-At. Mol. Opt.}\ }\textbf {\bibinfo {volume} {13}},\ \bibinfo {pages}
  {L461} (\bibinfo {year} {1980})}\BibitemShut {NoStop}%
\bibitem [{\citenamefont {Pendrill}\ and\ \citenamefont
  {Niemax}(1982)}]{pendrill1982}%
  \BibitemOpen
  \bibfield  {author} {\bibinfo {author} {\bibfnamefont {L.}~\bibnamefont
  {Pendrill}}\ and\ \bibinfo {author} {\bibfnamefont {K.}~\bibnamefont
  {Niemax}},\ }\href@noop {} {\bibfield  {journal} {\bibinfo  {journal} {J.
  Phys. B-At. Mol. Opt.}\ }\textbf {\bibinfo {volume} {15}},\ \bibinfo {pages}
  {L147} (\bibinfo {year} {1982})}\BibitemShut {NoStop}%
\bibitem [{\citenamefont {Boon}\ \emph {et~al.}(1999)\citenamefont {Boon},
  \citenamefont {Zekou}, \citenamefont {McGloin},\ and\ \citenamefont
  {Dunn}}]{boon1999}%
  \BibitemOpen
  \bibfield  {author} {\bibinfo {author} {\bibfnamefont {J.}~\bibnamefont
  {Boon}}, \bibinfo {author} {\bibfnamefont {E.}~\bibnamefont {Zekou}},
  \bibinfo {author} {\bibfnamefont {D.}~\bibnamefont {McGloin}}, \ and\
  \bibinfo {author} {\bibfnamefont {M.}~\bibnamefont {Dunn}},\ }\href@noop {}
  {\bibfield  {journal} {\bibinfo  {journal} {Phys. Rev. A}\ }\textbf {\bibinfo
  {volume} {59}},\ \bibinfo {pages} {4675} (\bibinfo {year}
  {1999})}\BibitemShut {NoStop}%
\bibitem [{\citenamefont {Urvoy}\ \emph {et~al.}(2013)\citenamefont {Urvoy},
  \citenamefont {Carr}, \citenamefont {Ritter}, \citenamefont {Adams},
  \citenamefont {Weatherill},\ and\ \citenamefont {L{\"o}w}}]{urvoy2013}%
  \BibitemOpen
  \bibfield  {author} {\bibinfo {author} {\bibfnamefont {A.}~\bibnamefont
  {Urvoy}}, \bibinfo {author} {\bibfnamefont {C.}~\bibnamefont {Carr}},
  \bibinfo {author} {\bibfnamefont {R.}~\bibnamefont {Ritter}}, \bibinfo
  {author} {\bibfnamefont {C.}~\bibnamefont {Adams}}, \bibinfo {author}
  {\bibfnamefont {K.}~\bibnamefont {Weatherill}}, \ and\ \bibinfo {author}
  {\bibfnamefont {R.}~\bibnamefont {L{\"o}w}},\ }\href@noop {} {\bibfield
  {journal} {\bibinfo  {journal} {J. Phys. B-At. Mol. Opt.}\ }\textbf {\bibinfo
  {volume} {46}},\ \bibinfo {pages} {245001} (\bibinfo {year}
  {2013})}\BibitemShut {NoStop}%
\bibitem [{\citenamefont {Arimondo}\ \emph {et~al.}(1977)\citenamefont
  {Arimondo}, \citenamefont {Inguscio},\ and\ \citenamefont
  {Violino}}]{arimondo1977}%
  \BibitemOpen
  \bibfield  {author} {\bibinfo {author} {\bibfnamefont {E.}~\bibnamefont
  {Arimondo}}, \bibinfo {author} {\bibfnamefont {M.}~\bibnamefont {Inguscio}},
  \ and\ \bibinfo {author} {\bibfnamefont {P.}~\bibnamefont {Violino}},\
  }\href@noop {} {\bibfield  {journal} {\bibinfo  {journal} {Rev. Mod. Phys.}\
  }\textbf {\bibinfo {volume} {49}},\ \bibinfo {pages} {31} (\bibinfo {year}
  {1977})}\BibitemShut {NoStop}%
\bibitem [{\citenamefont {Mohapatra}\ \emph {et~al.}(2007)\citenamefont
  {Mohapatra}, \citenamefont {Jackson},\ and\ \citenamefont
  {Adams}}]{mohapatra2007}%
  \BibitemOpen
  \bibfield  {author} {\bibinfo {author} {\bibfnamefont {A.}~\bibnamefont
  {Mohapatra}}, \bibinfo {author} {\bibfnamefont {T.}~\bibnamefont {Jackson}},
  \ and\ \bibinfo {author} {\bibfnamefont {C.}~\bibnamefont {Adams}},\
  }\href@noop {} {\bibfield  {journal} {\bibinfo  {journal} {Phys. Rev. Lett.}\
  }\textbf {\bibinfo {volume} {98}},\ \bibinfo {pages} {113003} (\bibinfo
  {year} {2007})}\BibitemShut {NoStop}%
\bibitem [{\citenamefont {Mack}\ \emph {et~al.}(2011)\citenamefont {Mack},
  \citenamefont {Karlewski}, \citenamefont {Hattermann}, \citenamefont
  {H{\"o}ckh}, \citenamefont {Jessen}, \citenamefont {Cano},\ and\
  \citenamefont {Fort{\'a}gh}}]{mack2011}%
  \BibitemOpen
  \bibfield  {author} {\bibinfo {author} {\bibfnamefont {M.}~\bibnamefont
  {Mack}}, \bibinfo {author} {\bibfnamefont {F.}~\bibnamefont {Karlewski}},
  \bibinfo {author} {\bibfnamefont {H.}~\bibnamefont {Hattermann}}, \bibinfo
  {author} {\bibfnamefont {S.}~\bibnamefont {H{\"o}ckh}}, \bibinfo {author}
  {\bibfnamefont {F.}~\bibnamefont {Jessen}}, \bibinfo {author} {\bibfnamefont
  {D.}~\bibnamefont {Cano}}, \ and\ \bibinfo {author} {\bibfnamefont
  {J.}~\bibnamefont {Fort{\'a}gh}},\ }\href@noop {} {\bibfield  {journal}
  {\bibinfo  {journal} {Phys. Rev. A}\ }\textbf {\bibinfo {volume} {83}},\
  \bibinfo {pages} {052515} (\bibinfo {year} {2011})}\BibitemShut {NoStop}%
\bibitem [{\citenamefont {Tauschinsky}\ \emph {et~al.}(2013)\citenamefont
  {Tauschinsky}, \citenamefont {Newell}, \citenamefont {van~den Heuvell},\ and\
  \citenamefont {Spreeuw}}]{tauschinsky2013}%
  \BibitemOpen
  \bibfield  {author} {\bibinfo {author} {\bibfnamefont {A.}~\bibnamefont
  {Tauschinsky}}, \bibinfo {author} {\bibfnamefont {R.}~\bibnamefont {Newell}},
  \bibinfo {author} {\bibfnamefont {H.~v.~L.}\ \bibnamefont {van~den Heuvell}},
  \ and\ \bibinfo {author} {\bibfnamefont {R.}~\bibnamefont {Spreeuw}},\
  }\href@noop {} {\bibfield  {journal} {\bibinfo  {journal} {Phys. Rev. A}\
  }\textbf {\bibinfo {volume} {87}},\ \bibinfo {pages} {042522} (\bibinfo
  {year} {2013})}\BibitemShut {NoStop}%
\bibitem [{\citenamefont {Sansonetti}(2008)}]{sansonetti2008wavelengths}%
  \BibitemOpen
  \bibfield  {author} {\bibinfo {author} {\bibfnamefont {J.}~\bibnamefont
  {Sansonetti}},\ }\href@noop {} {\bibfield  {journal} {\bibinfo  {journal} {J.
  Phys. Chem. Ref. Data}\ }\textbf {\bibinfo {volume} {37}},\ \bibinfo {pages}
  {7} (\bibinfo {year} {2008})}\BibitemShut {NoStop}%
\bibitem [{nis()}]{nist}%
  \BibitemOpen
  \href@noop {} {}\bibinfo {note} {NIST Standard Reference Database
  \#78}\BibitemShut {NoStop}%
\bibitem [{\citenamefont {Slanger}\ \emph {et~al.}(2000)\citenamefont
  {Slanger}, \citenamefont {Huestis}, \citenamefont {Cosby},\ and\
  \citenamefont {Osterbrock}}]{slanger2000accurate}%
  \BibitemOpen
  \bibfield  {author} {\bibinfo {author} {\bibfnamefont {T.}~\bibnamefont
  {Slanger}}, \bibinfo {author} {\bibfnamefont {D.}~\bibnamefont {Huestis}},
  \bibinfo {author} {\bibfnamefont {P.}~\bibnamefont {Cosby}}, \ and\ \bibinfo
  {author} {\bibfnamefont {D.}~\bibnamefont {Osterbrock}},\ }\href@noop {}
  {\bibfield  {journal} {\bibinfo  {journal} {J. Chem. Phys.}\ }\textbf
  {\bibinfo {volume} {113}},\ \bibinfo {pages} {8514} (\bibinfo {year}
  {2000})}\BibitemShut {NoStop}%
\bibitem [{\citenamefont {Civi{\v{s}}}\ \emph {et~al.}(2012)\citenamefont
  {Civi{\v{s}}}, \citenamefont {Ferus}, \citenamefont {Kubel{\'\i}k},
  \citenamefont {Jelinek},\ and\ \citenamefont
  {Chernov}}]{civivs2012potassium}%
  \BibitemOpen
  \bibfield  {author} {\bibinfo {author} {\bibfnamefont {S.}~\bibnamefont
  {Civi{\v{s}}}}, \bibinfo {author} {\bibfnamefont {M.}~\bibnamefont {Ferus}},
  \bibinfo {author} {\bibfnamefont {P.}~\bibnamefont {Kubel{\'\i}k}}, \bibinfo
  {author} {\bibfnamefont {P.}~\bibnamefont {Jelinek}}, \ and\ \bibinfo
  {author} {\bibfnamefont {V.}~\bibnamefont {Chernov}},\ }\href@noop {}
  {\bibfield  {journal} {\bibinfo  {journal} {Astron. Astrophys.}\ }\textbf
  {\bibinfo {volume} {541}},\ \bibinfo {pages} {A125} (\bibinfo {year}
  {2012})}\BibitemShut {NoStop}%
\bibitem [{\citenamefont {Abel}\ \emph {et~al.}(2009)\citenamefont {Abel},
  \citenamefont {Mohapatra}, \citenamefont {Bason}, \citenamefont {Pritchard},
  \citenamefont {Weatherill}, \citenamefont {Raitzsch},\ and\ \citenamefont
  {Adams}}]{abel2009}%
  \BibitemOpen
  \bibfield  {author} {\bibinfo {author} {\bibfnamefont {R.~P.}\ \bibnamefont
  {Abel}}, \bibinfo {author} {\bibfnamefont {A.~K.}\ \bibnamefont {Mohapatra}},
  \bibinfo {author} {\bibfnamefont {M.~G.}\ \bibnamefont {Bason}}, \bibinfo
  {author} {\bibfnamefont {J.~D.}\ \bibnamefont {Pritchard}}, \bibinfo {author}
  {\bibfnamefont {K.~J.}\ \bibnamefont {Weatherill}}, \bibinfo {author}
  {\bibfnamefont {U.}~\bibnamefont {Raitzsch}}, \ and\ \bibinfo {author}
  {\bibfnamefont {C.~S.}\ \bibnamefont {Adams}},\ }\href {\doibase
  http://dx.doi.org/10.1063/1.3086305} {\bibfield  {journal} {\bibinfo
  {journal} {Appl. Phys. Lett.}\ }\textbf {\bibinfo {volume} {94}},\ \bibinfo
  {eid} {071107} (\bibinfo {year} {2009})}\BibitemShut {NoStop}%
\end{thebibliography}%

\end{document}